\documentclass{article} 
\usepackage{iclr2025_conference,times}
\usepackage{graphicx}

\usepackage{amsmath,amsfonts,bm}

\def\eqref#1{equation~\ref{#1}}

\def\1{\bm{1}}

\DeclareMathAlphabet{\mathsfit}{\encodingdefault}{\sfdefault}{m}{sl}
\SetMathAlphabet{\mathsfit}{bold}{\encodingdefault}{\sfdefault}{bx}{n}

\usepackage{hyperref}
\usepackage{url}
\usepackage{acronym}

\title{When Openness Fails: Lessons from System Safety for Assessing Openness in AI}

\author{Tamara Paris\textsuperscript{\rm 1}, Shalaleh Rismani\textsuperscript{\rm 1}\\
\textsuperscript{\rm 1}McGill University, Montreal, Canada \\
\texttt{\{tamara.paris,shalaleh.rismani\}@mail.mcgill.ca}\\
}

\iclrfinalcopy 
\begin{document}

\acrodef{AI}{Artificial Intelligence}
\acrodef{OSI}{Open Source Initiative}
\acrodef{STAMP}{Systems-Theoretic Accident Model and Processes  }
\acrodef{STPA}{System Theoretic Process Analysis}
\acrodef{LLM}{Large Language Model}
\maketitle

\begin{abstract}
Most frameworks for assessing the openness of \ac{AI} systems use narrow criteria such as availability of data, model, code, documentation, and licensing terms. However, to evaluate whether the intended effects of openness---such as democratization and autonomy---are realized, we need a more holistic approach that considers the context of release: who will reuse the system, for what purposes, and under what conditions. To this end, we adapt five lessons from system safety that offer guidance on how openness can be evaluated at the system level.
\end{abstract}

\section{Introduction}

Derived from the notion of open source software, \textit{\ac{AI} openness} plays a major role in advancing practices of transparency, democratized access, and reusability in the \ac{AI} ecosystem. Although a wide variety of \ac{AI} models are labeled as ``open,''~\citep{examining_user_firendly_open_source_2023, opening_up_chatgpt_2023} the meaning of this designation is contested within the \ac{AI} community~\citep{open_for_business_big_tech_2023}. Recent efforts have proposed definitions and frameworks to characterize ``open'' or ``open source'' in AI~\citep{Openwashing_and_UE_Act_2024,Model_Openness_Framework_2024,OSI_Open_Source_AI_def}, often by adapting principles of open source software to distinctive characteristics of \ac{AI} systems. This adaptation is necessary because \ac{AI} systems differ from traditional software in composition (e.g., unlike source code, model parameters are not human-readable)~\citep{towards_a_framework_for_Openness_2024}, can be released and made accessible using various approaches~\citep{Gradient_of_generative_ai_release_2023, beyond_release_access_considerations_2025} and raise unique challenges, notably for interpretability and oversight (e.g., the ``black box'' problem).

In high-stakes domains, such as scientific research and policymaking, openness is often seen as essential for advancing transparency, accountability, and collective progress~\citep{using_proprietary_language_models_explicit_justification_2024}. Yet whether these benefits are realized depends on the conditions under which openness is pursued. Existing frameworks for evaluating whether openness is achieved often overlook contextual factors of an \ac{AI} system release, such as the target audience or legal environment, even though these factors strongly shape whether aforementioned benefits are realized. When such contextual dependencies are ignored, making an \ac{AI} system ``open'' can fail to deliver its intended benefits or even introduce new risks. We refer to these cases as “failures of openness” (Sec.~\ref{subsec:failure_openness}) and explain how a system safety perspective could help us in framing/understanding these ``failures'' at the system-level. (Sec.~\ref{sec:parallel_system_safety}). Lastly, through a fictional scenario, we will detail how openness could be assessed differently, drawing inspiration from five of the system safety lessons described by~\citet{System_safety_and_artificial_intelligence_2022} (Sec.~\ref{sec:system_safety_insights}).

\subsection{Failure of AI Openness}
\label{subsec:failure_openness}

In a recent work, ~\citet{Opening_the_scope_of_openness_in_AI_2025} identify three ways to frame openness for \ac{AI} systems: 1) \textit{properties} of the system and its release method (e.g., license terms, components availability), 2) \textit{afforded actions} (e.g., use, modify, share), and  3) \textit{desired effects} (e.g., autonomy, transparency). Current frameworks and definitions of openness in \ac{AI} focus on only one of these framings. For example, ~\citet{OSI_Open_Source_AI_def} defines openness through \textit{afforded actions} (use, study, modify and share the \ac{AI} system) and ~\citet{Openwashing_and_UE_Act_2024} assess concrete \textit{properties} of the release method (e.g., model card availability, license type) to evaluate \ac{AI} model openness.

However, the \textit{properties} of an \ac{AI} system and its release method -- factors often controlled by AI developers/organizations -- do not always lead to the intended \textit{afforded actions} and \textit{desired effects}. These elements also depend on the broader release context and power dynamics. For example, model weights may be released under a license that allows modification, but if the computing power required to fine-tune these weights is very high, most individuals would be unable to modify them.
However, accessibility of computing resources depends on external factors such as chip price or export control restrictions, which are beyond the direct control of \ac{AI} developers. In this case, the model's release method fails to enable an \textit{afforded action} commonly associated with \ac{AI} openness (i.e., modify the model), although it meets the expected \textit{properties} of an open model. Likewise, some \textit{afforded actions} may not ultimately lead to the \textit{desired effects} of openness.
We refer to such issues as \textit{failures of openness} and aim to address the lack of means to detect these failures before the model release. 

\subsection{The Parallel between System Safety and AI Openness}
\label{sec:parallel_system_safety}

Assessing how and why a complex system may fail to reach a desired outcome (i.e., safe operation) is also a challenge faced in the field of safety. 
System safety is an established engineering discipline that treats safety as an emergent property of the whole system rather than its individual components. It focuses on preventing losses such as injury, death, and damage to property in complex systems used in domains such as aerospace and automotive~\citep{Leveson_2012}. Unlike earlier accident models that traced failures to chains of component breakdowns or operator errors, system safety emphasizes how safety depends on system interactions (e.g., the connection between the car brake and the self-driving software), the enforcement of appropriate constraints (i.e., required braking distance), and functioning feedback mechanisms (i.e., dashboard alerts)~\citep{Leveson_2012}. Recently, lessons from system safety have been applied to responsible \ac{AI} development \cite{System_safety_and_artificial_intelligence_2022, Rismani2024-qc}. 

We argue that the system safety perspective is useful for understanding \ac{AI} openness. 
Failures of openness---such as when transparency, adaptability, or reusability are not achieved---can be understood as a form of loss. Current frameworks of assessing openness often treat it as a fixed property of individual \ac{AI} system components, such as whether model weights or documentation are publicly accessible. However, we argue that openness should be evaluated as a dynamic, system-level condition that is shaped by the distribution of control (e.g., who decides to release what part of the model), the presence or absence of feedback mechanisms (e.g., error reporting channels), and the assumptions embedded in sociotechnical interactions (e.g., the expertise of the potential users). 
By adapting five lessons from system safety to \ac{AI}  openness, we lay the groundwork for more rigorous and context-sensitive approaches to assessing openness in \ac{AI}. 

\section{System Safety Insights for AI Openness}
\label{sec:system_safety_insights}

\subsection{Example Scenario}

\begin{figure}[h]
    \centering
    \includegraphics[width=1\linewidth]{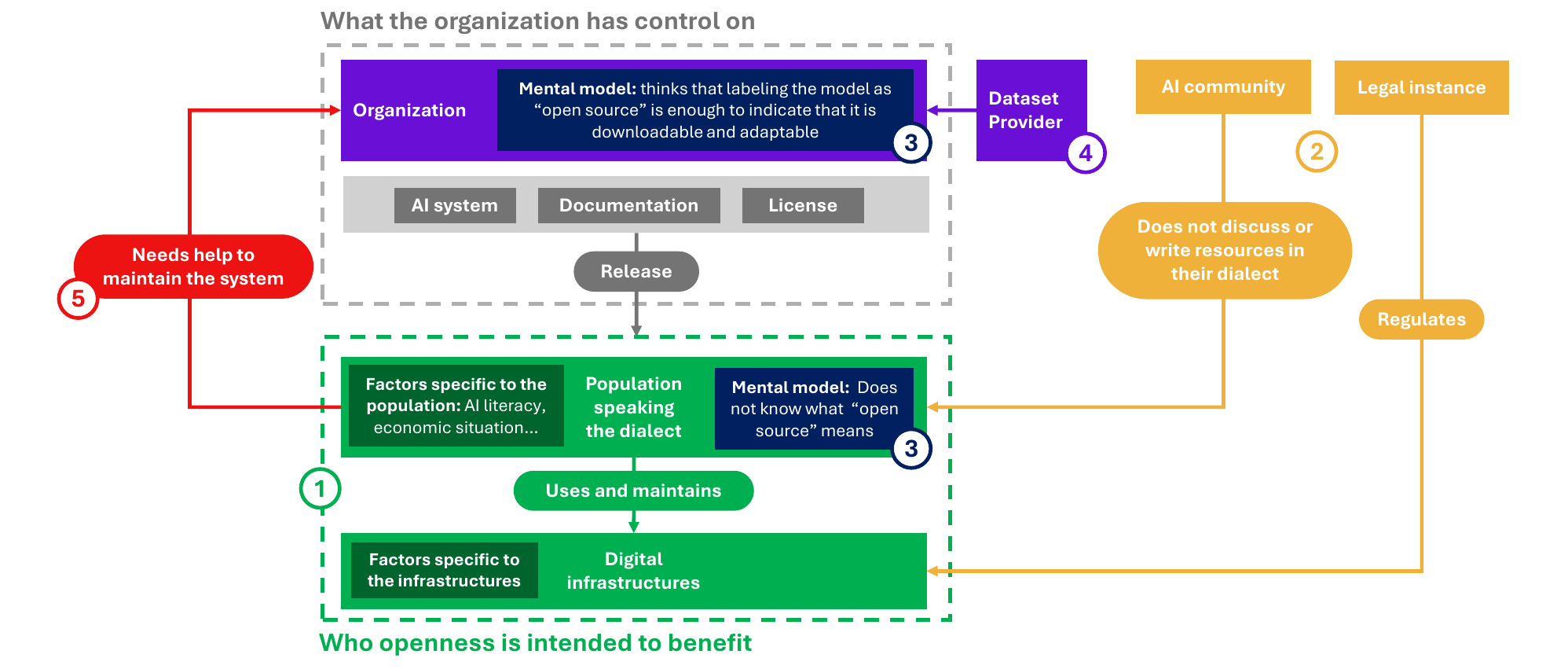}
    \caption{Illustration of the fictional example scenario. Each part of the diagram is associated with the lesson number it corresponds to.}
    \label{fig:diagram}
\end{figure}

\begin{figure}[h]
    \centering
    \includegraphics[width=1\linewidth]{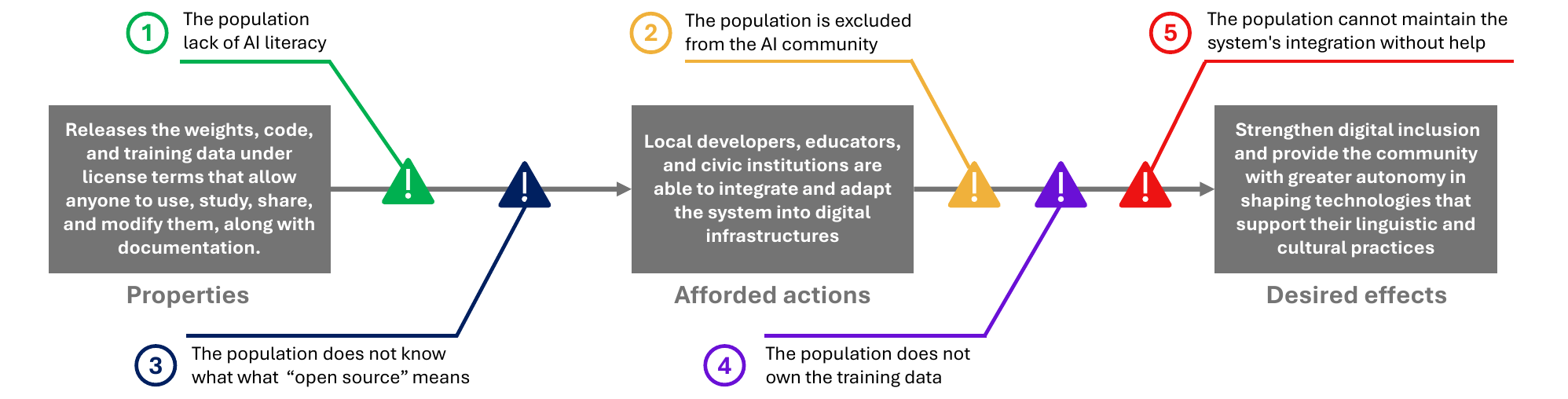}
    \caption{Examples of hazards that can lead to a \textit{failure of openness} for each of the five lessons. The lessons are indicated by the numbers and colors.}
    \label{fig:failures}
\end{figure}

To highlight the relevance of system safety insights for assessing openness in \ac{AI}, we illustrate them through a fictional example scenario: an organization that releases the weights, code, and training data of an \ac{LLM} specialized in one specific dialect spoken by a minority community. 
The components of the \ac{AI} system are released under license terms that allow anyone to use, study, share, and modify them, along with documentation. 
The organization wants to open the model to encourage local developers, educators, and civic institutions to integrate and adapt the system into the digital infrastructure used by the population speaking the dialect. The hope is that this process will strengthen digital inclusion and provide the community with greater autonomy in shaping technologies that support their linguistic practices. In the following section, we elaborate on five lessons for assessing \ac{AI} openness drawn from system safety and apply them to this scenario, illustrated in Figure ~\ref{fig:diagram}. Figure~\ref{fig:failures} illustrates examples of hazards (i.e., potential sources of harm) that can lead to failures of openness across the \textit{properties}, \textit{afforded actions}, and \textit{desired effects} expected in this fictional scenario. 

\subsection{Lessons from system safety}

\textbf{Lesson 1:} 
\textit{The openness of an AI system should be assessed in its deployment context, by taking into account the targeted audience and environment}. 

This lesson builds on the core idea that safety is not just about whether each part of a system functions as intended, but about how people, technologies, and institutions interact. 
From this system-theoretic perspective, \textit{afforded actions} and \textit{desired effects} of openness not only depend on the license terms and the availability of individual components, but also on the capacities and constraints of the people and institutions involved.
In our fictional scenario, the openness of the \ac{LLM} specialized in dialect should be evaluated by whether it enables adaptability and provides autonomy to the community that speaks it. In particular, factors specific to this population (e.g., its \ac{AI} literacy, its economic situation) affect its ability to adapt the model to their needs.

\textbf{Lesson 2:} \textit{Properly assessing openness requires analyzing the sociotechnical constraints that shape how technology, people, and institutions interact}. 

System safety highlights that simple causal chains are insufficient to describe the process leading to an accident. Similarly, failures of openness in \ac{AI} are often not the result of a chain of linear causes, but stem from complex sociotechnical constraints, such as cultural practices, power dynamics, legal frameworks and political incentives. 
In our scenario, the inability of the target population to achieve autonomy by adapting the open model cannot be explained solely by causes such as limited economic resources or low \ac{AI} literacy.
For example, the same population may be subject to other constraints, such as restrictive regulations on its digital infrastructures. It may also be systematically excluded from the mainstream \ac{AI} community. This may happen if resources and discussions are never in their dialect, or if structural discrimination blocks their access. In such cases, there will always be a power imbalance between the mainstream \ac{AI} community and this population. Identifying this type of imbalance is crucial for evaluating whether openness efforts are truly achieving their desired effects.

\textbf{Lesson 3:}  \textit{The targeted population may have a different mental model of what openness is than the developers. Reaching openness depends on holding adequate/complete mental models.}

From a system safety perspective, accidents often occur because different actors, such as operators, engineers, or automated systems, hold incomplete or mismatched understandings of how the system works. These divergent and inaccurate mental models can lead to unsafe decisions even when system components are functioning correctly. A similar issue arises with openness in \ac{AI}. People who could benefit from the openness of a model have their own beliefs, assumptions, and knowledge about how the system works or how to use it. Their mental model may therefore affect their ability to carry out actions that are technically possible for them but out of their purview.  
In our fictional scenario, the target population may not know what ``open source'' means, or misunderstand it. In this situation, even if the organization markets the model as ``open source,'' the population would probably not know that they can download the parameters and modify them.  

\textbf{Lesson 4:}  \textit{Organizational dependencies and dynamics can block openness at the time of release. Therefore, they should be considered from the beginning of the AI lifecycle.}

~\citet{System_safety_and_artificial_intelligence_2022} describes how outsourcing a part of the \ac{AI} development can lead to more elusive safety goals and less ability to regulate the system. \ac{AI} openness faces a similar problem: the development of an \ac{AI} system may involve multiple organizations operating at different stages of the \ac{AI} lifecycle, and thus the organization releasing the \ac{AI} system may not have full control over the openness of each individual component. In our scenario, the organization may have trained the model on a dataset owned by neither the organization nor the community that speaks the dialect. If later versions of the dataset are not released openly, or if the population cannot remove data from it, the lack of control over the training data can undermine the openness of the system. Such dependencies highlight how organizational choices throughout the lifecycle shape whether openness goals can actually be met.

\textbf{Lesson 5:} 
\textit{Openness is not a fixed property of an AI system, but one that evolves over time. Therefore, assessment must take into account the efforts put in place to maintain the openness over time.} 

System safety emphasizes that systems can drift into states that carry higher risks, even if they were initially safe. Continuous monitoring and corrective mechanisms are required to maintain safety over time. Applying this perspective to openness highlights the importance of treating openness as a dynamic condition rather than a one-time attribute.
In our fictional scenario, institutions that integrate the dialect-specific \ac{LLM} into their infrastructures may initially succeed in adapting it to local needs. However, if these institutions lack the financial and human resources to maintain the \ac{AI} system's integration without constant help from the organization that developed it, it greatly reduces autonomy beyond the temporary ability to adapt the model to their needs. Assessing openness at release must therefore consider whether the population can sustain the use and adaptation of the model over the long term, ensuring that the desired effects of openness extend beyond the moment of release.

\bibliography{iclr2025_conference}
\bibliographystyle{iclr2025_conference}

\end{document}